\documentclass[%
 reprint,
 amsmath,amssymb,superscriptaddress,
 aps,
]{revtex4-2}

\usepackage{xcolor}

\usepackage{graphicx}
\usepackage{dcolumn}
\usepackage{bm}
\usepackage{physics}

\begin{document}
\title{Magnetic resonance study of rare-earth titanates}

\author{A. Najev}
\affiliation{Department of Physics, Faculty of Science, University of Zagreb, Bijeni\v cka 32, HR-10000 Zagreb, Croatia}
\affiliation{School of Physics and Astronomy, University of Minnesota, Minneapolis, MN 55455, U.S.A.}
\author{S. Hameed}
\altaffiliation{Present address: Max-Planck-Institut for Solid State Research, Heisenbergstraße 1, D-70569 Stuttgart, Germany}
\affiliation{School of Physics and Astronomy, University of Minnesota, Minneapolis, MN 55455, U.S.A.}

\author{A. Alfonsov}
\affiliation{Leibniz Institute for Solid State and Materials Research IFW Dresden, 01069 Dresden, Germany}
\author{J. Joe}
\affiliation{School of Physics and Astronomy, University of Minnesota, Minneapolis, MN 55455, U.S.A.}
\author{V. Kataev}
\affiliation{Leibniz Institute for Solid State and Materials Research IFW Dresden, 01069 Dresden, Germany}
\author{M. Greven}
\affiliation{School of Physics and Astronomy, University of Minnesota, Minneapolis, MN 55455, U.S.A.}
\author{M. Požek}
\affiliation{Department of Physics, Faculty of Science, University of Zagreb, Bijeni\v cka 32, HR-10000 Zagreb, Croatia}
\author{D. Pelc}
\email{correspondence to: dpelc@phy.hr}
\affiliation{Department of Physics, Faculty of Science, University of Zagreb, Bijeni\v cka 32, HR-10000 Zagreb, Croatia}
\affiliation{School of Physics and Astronomy, University of Minnesota, Minneapolis, MN 55455, U.S.A.}

\begin{abstract} 
We present a nuclear magnetic resonance (NMR) and electron spin resonance (ESR) study of rare-earth titanates derived from the spin-1/2 Mott insulator YTiO$_3$. Measurements of single-crystalline samples of (Y,Ca,La)TiO$_3$ in a wide range of isovalent substitution (La) and hole doping (Ca) reveal several unusual features in the paramagnetic state of these materials. $^{89}$Y NMR demonstrates a clear discrepancy between the static and dynamic local magnetic susceptibilities, with deviations from Curie-Weiss behavior far above the Curie temperature $T_C$. No significant changes are observed close to $T_C$, but a suppression of fluctuations is detected in the NMR spin-lattice relaxation time at temperatures of about $3\times T_C$. Additionally, the nuclear spin-spin relaxation rate shows an unusual peak in dependence on temperature for all samples. ESR of the unpaired Ti electron shows broad resonance lines at all temperatures and substitution/doping levels, which we find to be caused by short electronic spin-lattice relaxation times. We model the relaxation as an Orbach process that involves a low-lying electronic excited state, which enables the determination of the excited-state gap from the temperature dependence of the ESR linewidths. We ascribe the small gap to Jahn-Teller splitting of the two lower Ti $t_{2g}$ orbitals. The value of the gap closely follows $T_C$ and is consistent with the temperatures at which deviations from Curie-Weiss fluctuations are observed in NMR. These results provide insight into the interplay between orbital and spin degrees of freedom in rare-earth titanates and indicate that full orbital degeneracy lifting is associated with ferromagnetic order.  
\end{abstract}
\maketitle

\section{Introduction}

\begin{figure}
\includegraphics[width=0.48\textwidth]{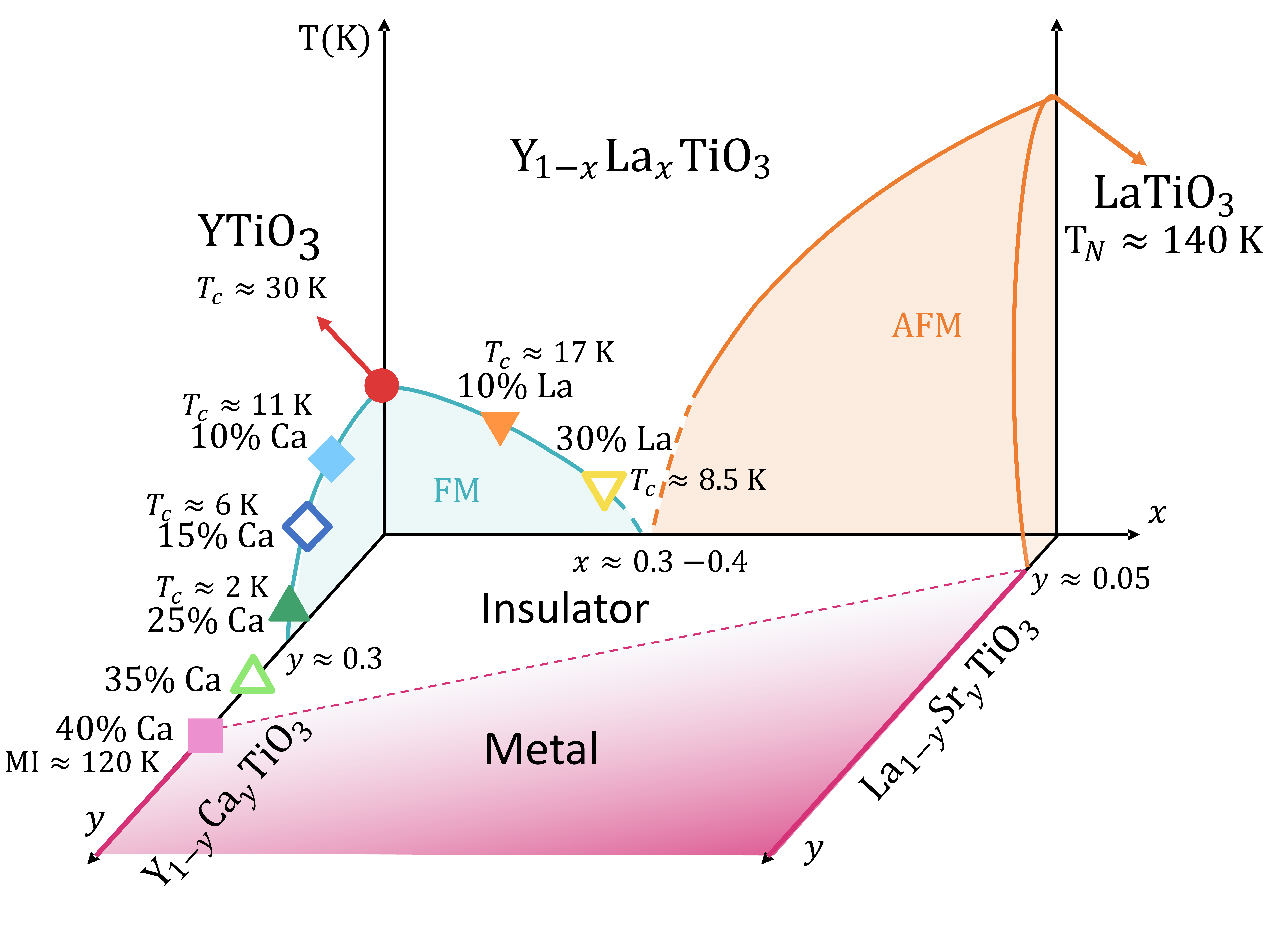}
\caption{\label{fig:fdija} Schematic phase diagram of the R$_{1-x}$A$_x$TiO$_3$ (R= Y, La, A= Ca, Sr) system. Compounds studied with NMR and ESR in this work are marked with symbols.}
\end{figure}

Among transition metal oxides with strong electron-electron correlations, the rare-earth (RE) titanates exhibit perhaps the simplest electronic structure, which makes them model systems for the investigation of spin-lattice coupling, orbital degeneracy effects, and metal-insulator transitions \cite{tokSci}. The electronic and magnetic properties of these materials stem from unpaired electrons in the lower $t_{2g}$ orbitals and GdFeO$_3$-type octahedral distortions away from the ideal cubic perovskite structure \cite{revijalac}. Depending on the size of the rare-earth (R) ion in RTiO$_3$, the TiO$_6$ octahedra rotate and tilt such that the Ti-O-Ti bond angle deviates more strongly from 180$^\circ$ for smaller R ions \cite{struktura}. These structural distortions influence the orbital overlap and the superexchange interaction between the unpaired Ti spins. For R=\{Yb, Y, Gd\} the distortion is relatively large, and a predominant ferromagnetic (FM) ground state appears, with spins approximately along [001] (the $c$ axis). Ferromagnetism in oxides with strong electronic correlations is unusual, and its origin in the titanates is the subject of ongoing debate. In contrast, for larger RE ions R=\{Sm, Nd, Pr, Ce, La\}, the magnetic order is predominantly antiferromagnetic (AFM) G-type along [100] (the $a$ axis). Isovalent substitution of Y with La enables nearly continuous tuning of the average RE ion radius, and leads to an evolution from FM to AFM order with increasing La concentration, concomitant with a decrease of the GdFeO$_3$-type distortion. Similar effects have recently been found with applied uniaxial stress, which also modifies the octahedral distortions \cite{knafo, Zhaoorbliq, najev}. On the other hand, hole doping in Y$_{1-y}$Ca$_y$TiO$_3$ and La$_{1-y}$Sr$_y$TiO$_3$ destroys the long-range magnetic order and leads to an insulator-metal transition (Fig.\,\ref{fig:fdija}) \cite{cadop, cadop2, tokura, HameedCa, HameedCaErr}. 


The rare-earth titanates have been the subject of resurgent interest. As prototypical three-dimensional low-spin Mott insulators with a doping-controlled metal-insulator transition, these materials represent well-defined model systems for the study of the underlying mechanisms that govern properties such as superconductivity and colossal magnetoresistance in other oxides. It has been shown that the Mott transition in Ca-doped YTiO$_3$, which occurs far from half-filling, likely involves phase separation, with implications for a wide range of systems \cite{HameedCa, HameedCaErr}. The nature of the magnetic transitions and their relation to orbital physics has also been discussed extensively. Recent work has determined the size of the saturated magnetic moment in YTiO$_3$, demonstrated that the orbital moment is fully quenched, and shown that both the size of the moment and the magnetic volume fraction decrease with La substitution \cite{HameedLa}. However, it is still unclear if orbital degeneracy is present in the paramagnetic and AFM-ordered phases, and if there exists an ``orbital liquid" ground state \cite{Zhaoorbliq}. Moreover, the paramagnetic state is likely unconventional, with significant anomalous magnetic contributions to thermal expansion and specific heat well above $T_C$ in YTiO$_3$ \cite{knafo}. Deviations from a Curie-Weiss law have been detected in magnetisation measurements at similar temperatures \cite{kovaleva1,cheng}, along with a subtle optical phonon spectral weight redistribution which could be caused by slight lattice distortions that change the local symmetry \cite{kovaleva2}. Additionally, a non-equilibrium FM state at temperatures reaching about $2.5\,T_C$ in YTiO$_3$ is created by optical stabilization of the oxygen rotation modes \cite{disa}. All of this points to the existence of a precursor to the FM state, with an energy scale between $2$ and $3\,T_C$ in YTiO$_3$, whose nature and relation to the orbital physics remains unknown. In an effort to shed light on this question, we employ two magnetic resonance techniques -- nuclear magnetic resonance (NMR) and electron spin resonance (ESR) -- to obtain deeper insight into the local magnetism of the paramagnetic state of the (Y,La,Ca)TiO$_3$ system.

Previous local-probe investigations of rare-earth titanates were mostly carried out in the magnetically ordered phases at temperatures below $T_C$ and $T_N$, including NMR and muon spin relaxation  \cite{HameedLa, Voufack:gq5011, neutYTO, polneutr, magnetoelast, bing, haverkort, ulrich2002}. In addition, most NMR studies of RE titanates have focused on the nature and extent of orbital order in stoichiometric compounds \cite{NMRKiyama, NMRKiyama2, NMRItoh, NMRItoh2,NMRT1,furu1996}. One exception is a Ti NMR study of doped La$_{1-y}$Sr$_y$TiO$_3$, which exhibits a insulator-metal transition at very low Sr concentrations, where a significant decrease of the linewidth and relaxation rates is observed in the metallic phase \cite{NMRFuru}. However, the insulating part of the (Y,La,Ca)TiO$_3$ phase diagram remains unexplored. ESR is complementary to NMR and directly probes the unpaired electrons responsible for the magnetism. Previous ESR work was also limited to low temperatures and stoichiometric compounds; a FM resonance was detected in the ordered phase of YTiO$_3$, consistent with an easy-plane-type anisotropy that could potentially be traced above $T_C$ \cite{esrsubmil}. 

In this paper, we present NMR and X-band ESR measurements of (Y,La,Ca)TiO$_3$ for a wide range of doping/substitution levels, at temperatures above the magnetic transition (Fig.\,\ref{fig:fdija}). We systematically explore the still poorly understood paramagnetic phase, with the central result that low-lying orbital-excited states are generically present and important for understanding the magnetic precursor effects in these materials.

This paper is organized as follows: Section II features experimental details; Sections III and IV present the results of the NMR and ESR experiments, respectively; in Section V, we discuss and summarize our results.

\section{Experimental}

The NMR and ESR experiments were performed on well-characterized single crystals of YTiO$_3$, Y$_{1-x}$La$_{x}$TiO$_3$ ($x=$0.1, 0.3) and Y$_{1-y}$Ca$_y$TiO$_3$ ($y=$0.1, 0.25, 0.35, 0.4), marked with symbols on the schematic phase diagram shown in Fig.\,\ref{fig:fdija}. The crystals were grown using the traveling-solvent floating-zone technique \cite{hameedgrowth}. Characterization included chemical composition analysis, magnetometry, charge transport, neutron scattering, x-ray absorption spectroscopy and x-ray magnetic circular dichroism, as shown elsewhere \cite{hameedgrowth, HameedCa, HameedCaErr, HameedLa}. Temperatures of the magnetic phase transitions measured in near-zero external field are marked in Fig.\,\ref{fig:fdija}; all samples are either in the ferromagnetic or the paramagnetic region of the phase diagram.

The $^{89}$Y NMR measurements were performed in a high-homogeneity 12\,T superconducting magnet (Oxford Instruments) using commercial (Tecmag Apollo and Redstone) broadband spectrometers and Tomco power amplifiers. The crystalline $c$-axis was parallel to the field for all samples. NMR spectra were obtained with the conventional Hahn echo sequence $\pi/2-\tau-\pi$. $^{89}$Y NMR frequencies were below 25\,MHz, so the high-Q coils exhibited pronounced ringing after the radiofrequency pulses, and $\tau$ needed to be longer than $\sim80$\,$\mu$s to avoid ringing artefacts. The $\pi/2$ pulse length was 8\,-\,12\,$\mu$s, depending on the sample and coil size. The spin-lattice relaxation time $T_1$ was measured using a saturation recovery sequence $\pi/2-\text{Delay}-\pi/2-\tau-\pi$ with a logarithmic variation of the delay time. The spin-spin relaxation time $T_2$ was determined from echo decay; we used the standard Hahn echo sequence with varying delay $\tau$ after the excitation pulse. Long optimal pulses and $\tau$ times limited our $T_2$ measurements to a minimum echo time of about 150\,$\mu$s. We used conventional anti-ringing pulse phase rotations for all measurement sequences \cite{nutsnboltsNMR}. 

\begin{figure*}
\includegraphics[width=175mm]{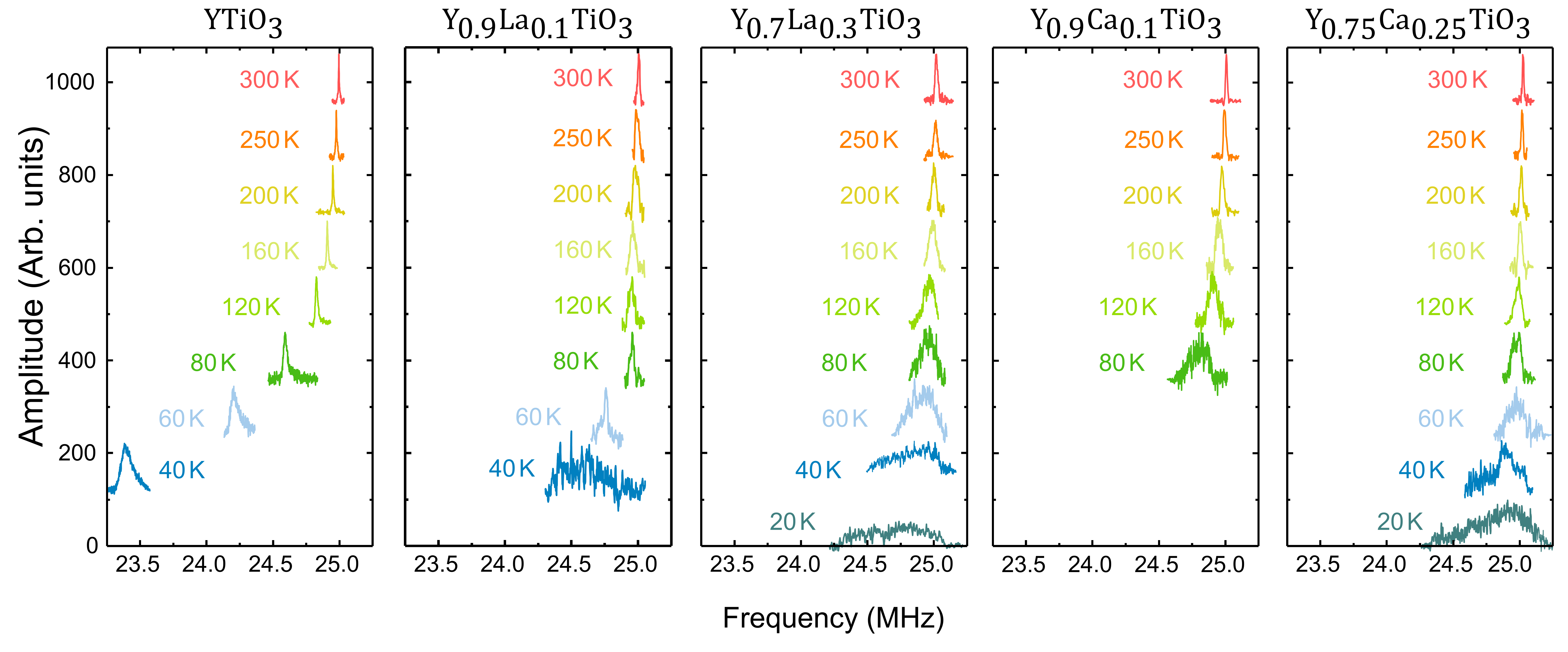}
\caption{\label{fig:spec} Temperature dependence of $^{89}$Y NMR spectra in a 12\,T field for different samples. The individual spectra are scaled and shifted vertically for comparison. The stoichiometric compound YTiO$_3$ shows a large shift above the Curie temperature, whereas doping decreases the shift significantly and increases the low-temperature linewidths.}
\end{figure*}

ESR measurements were performed with a commercial Bruker X-band spectrometer equipped with a He-gas-flow cryostat that enabled measurements in a temperature range of 3.5\,-\,300\,K. Samples were mounted on an automated goniometer with the axis of rotation perpendicular to the field direction. ESR measurements were continuous-wave with a fixed frequency of about 9.5\,GHz. The magnetic field was swept up to 0.9\,T and modulated with a 10\,G amplitude, and all ESR spectra are presented as the magnetic field derivative of the absorbed microwave power $dP(H)/dH$. The excitation power was kept constant at 63\,$\mu$W for all measurements. Because of high absorption, ESR signals closer to the FM transition temperature needed to be attenuated up to 45\,dB and the receiver gain needed to be adjusted.


\section{NMR results}

Our NMR experiments focused on $^{89}$Y; a search for Ti NMR signals in the paramagnetic phase was unsuccessful, likely due to the strongly fluctuating Ti spins that lead to very fast nuclear relaxation rates \cite{NMRFuru}. $^{89}$Y is a spin-1/2 nucleus, and thus insensitive to electric field gradients, which significantly simplifies the analysis of the results. The Y ion has eight approximately equidistant Ti nearest neighbors. Therefore, assuming that only the Ti atoms show appreciable magnetization, the $^{89}$Y NMR form factor is peaked at $q = 0$, \textit{i.e.}, $^{89}$Y NMR is only sensitive to FM correlations on Ti.

\begin{figure}
\includegraphics[width=0.48\textwidth]{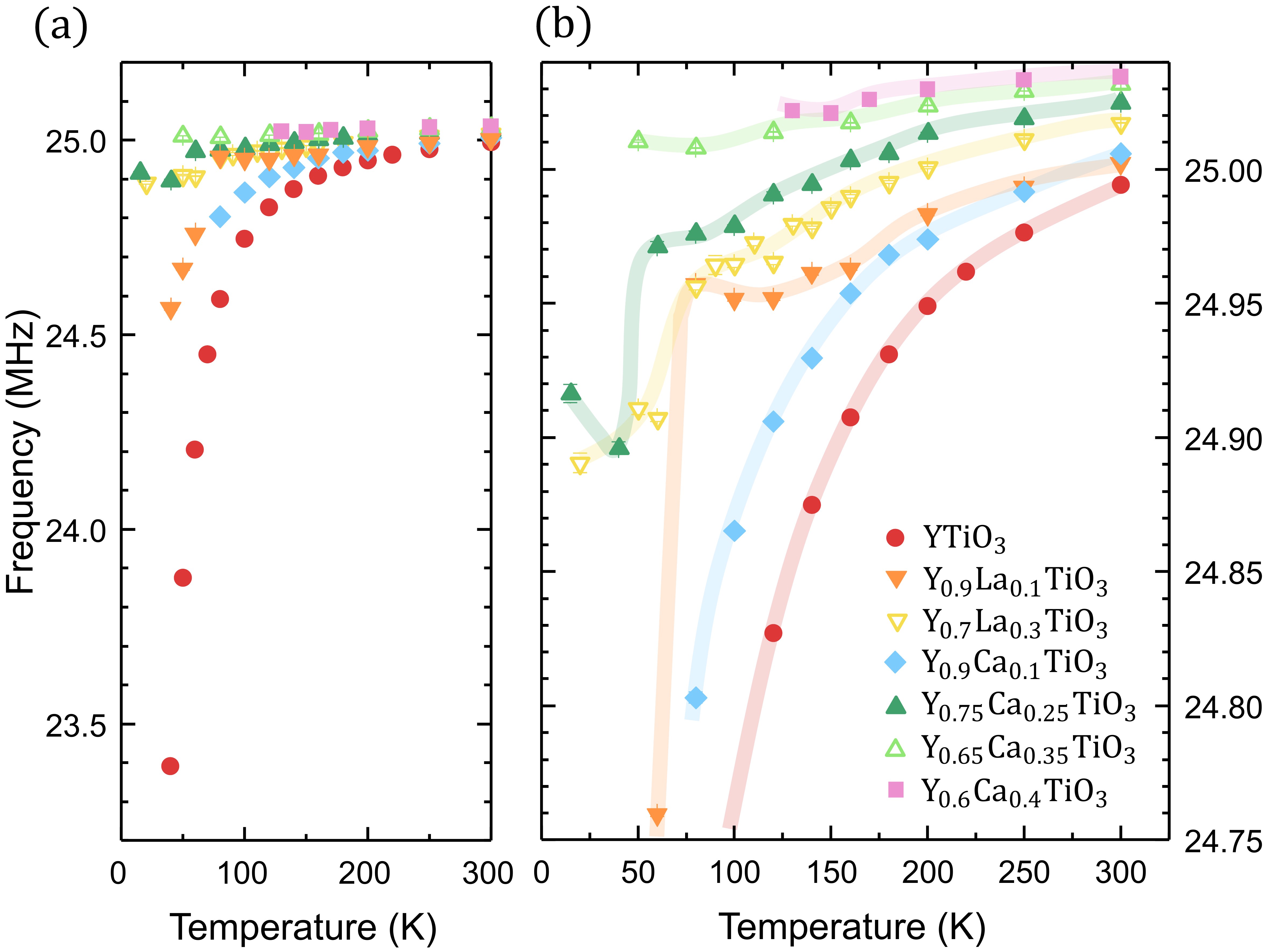}
\caption{\label{fig:freq} Temperature dependence of $^{89}$Y spectra peak frequencies for YTiO$_3$, Y$_{1-x}$La$_{x}$TiO$_3$ ($x=$0.1, 0.3) and Y$_{1-x}$Ca$_x$TiO$_3$ ($x=$0.1, 0.25, 0.35, 0.4), obtained from Gaussian fits. (a) full frequency range; (b) expanded view. All lines are guides to the eye.}
\end{figure}

Representative $^{89}$Y NMR spectra are shown in Fig.\,\ref{fig:spec}. In NMR, the shift of the resonant peak position compared to a reference frequency is directly proportional to the local static spin susceptibility. The constant of proportionality is referred to as the hyperfine coupling, and its sign and value depend on the nature of the interactions between nuclear and electronic spins and the details of the Y atomic configuration. Upon cooling from room temperature to $T_C=30$\,K, the $^{89}$Y line in stoichiometric YTiO$_3$ is seen to shift substantially toward lower frequencies. This observation is in agreement with published results for YTiO$_3$ \cite{NMRItoh2,furu1996,NMRT1} and in line with expectations for a material with FM fluctuations and a negative hyperfine coupling.

The shift is less prominent in the La-substituted and Ca-doped samples. However, we find that the lines broaden significantly at low temperatures. The broadening is most likely inhomogeneous, given that the $T_2$ values are not short enough to explain the linewidths (see below). This implies that a spatial distribution of static local fields appears as the FM phase is approached. As a measure of the average local susceptibility that can be compared among samples, we extract peak frequencies from Gaussian fits of the $^{89}$Y spectra, shown in Fig.\,\ref{fig:freq}. The substitution of both La and Ca significantly suppresses the shifts at low temperatures, which is qualitatively consistent with the decrease of the Ti ordered moment observed in neutron diffraction, x-ray magnetic circular dichroism, magnetization and $\mu$SR \cite{HameedLa}. The broadening of the resonance lines makes a precise determination of the peak frequency difficult. Moreover, a significant shortening of the $T_2$ relaxation time occurs in this temperature range. This unexpected behavior will be discussed below.

\begin{figure}
\includegraphics[width=0.48\textwidth]{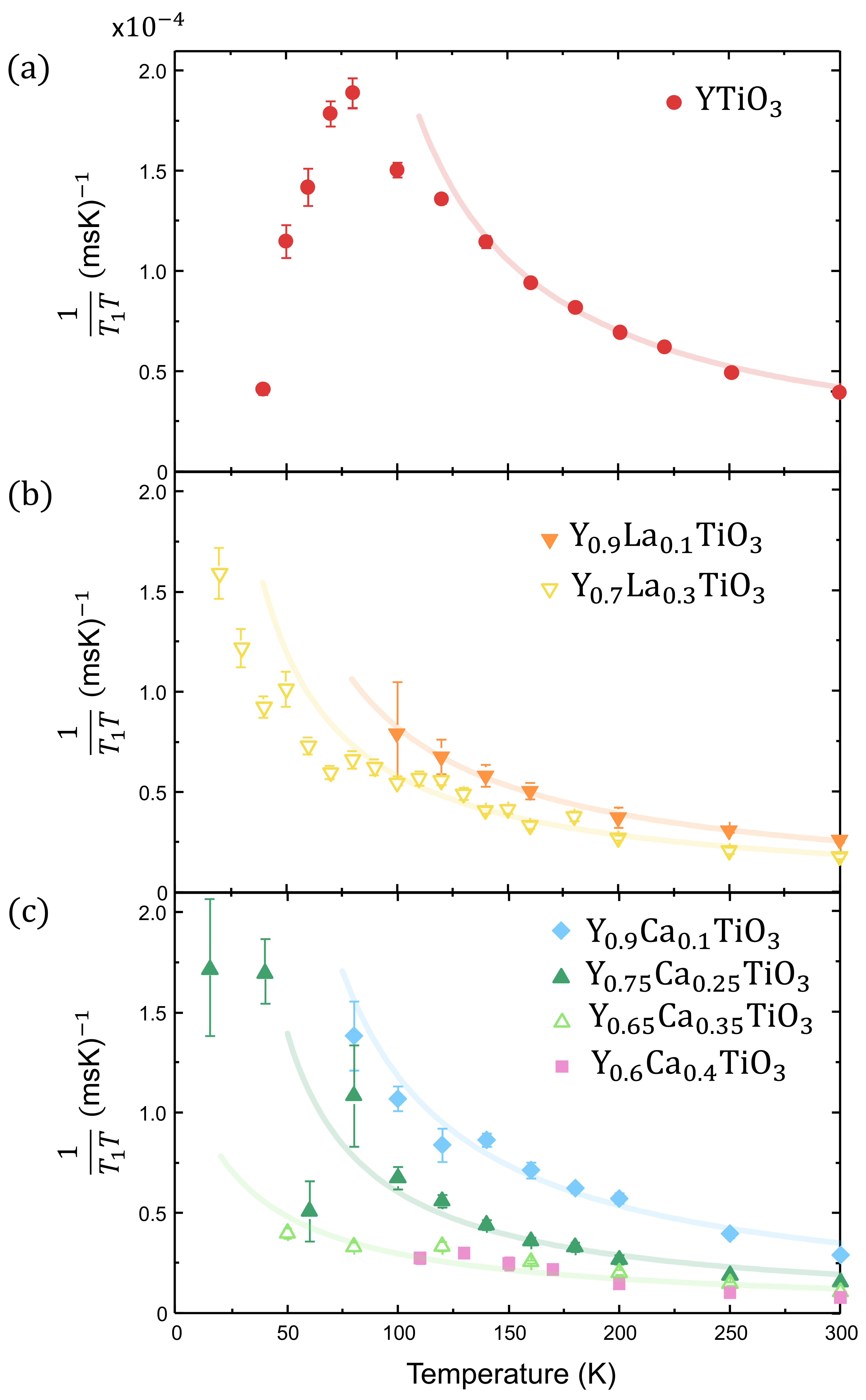}
\caption{\label{fig:T1} Temperature dependence of the dynamic spin susceptibility $1/(T_1T)$ for the $^{89}$Y nucleus. The lines are best fits to the Curie-Weiss form at high temperatures (for fit range, see Table\,\ref{tab:table1}). (a) A clear downward deviation from Curie-Weiss behavior is seen in the undoped parent compound YTiO$_3$ below $\sim120\,$K. Similar, but weaker effects are found in (b) La-substituted and (c) Ca-doped samples.}
\end{figure}

\begin{figure}
\includegraphics[width=0.45\textwidth]{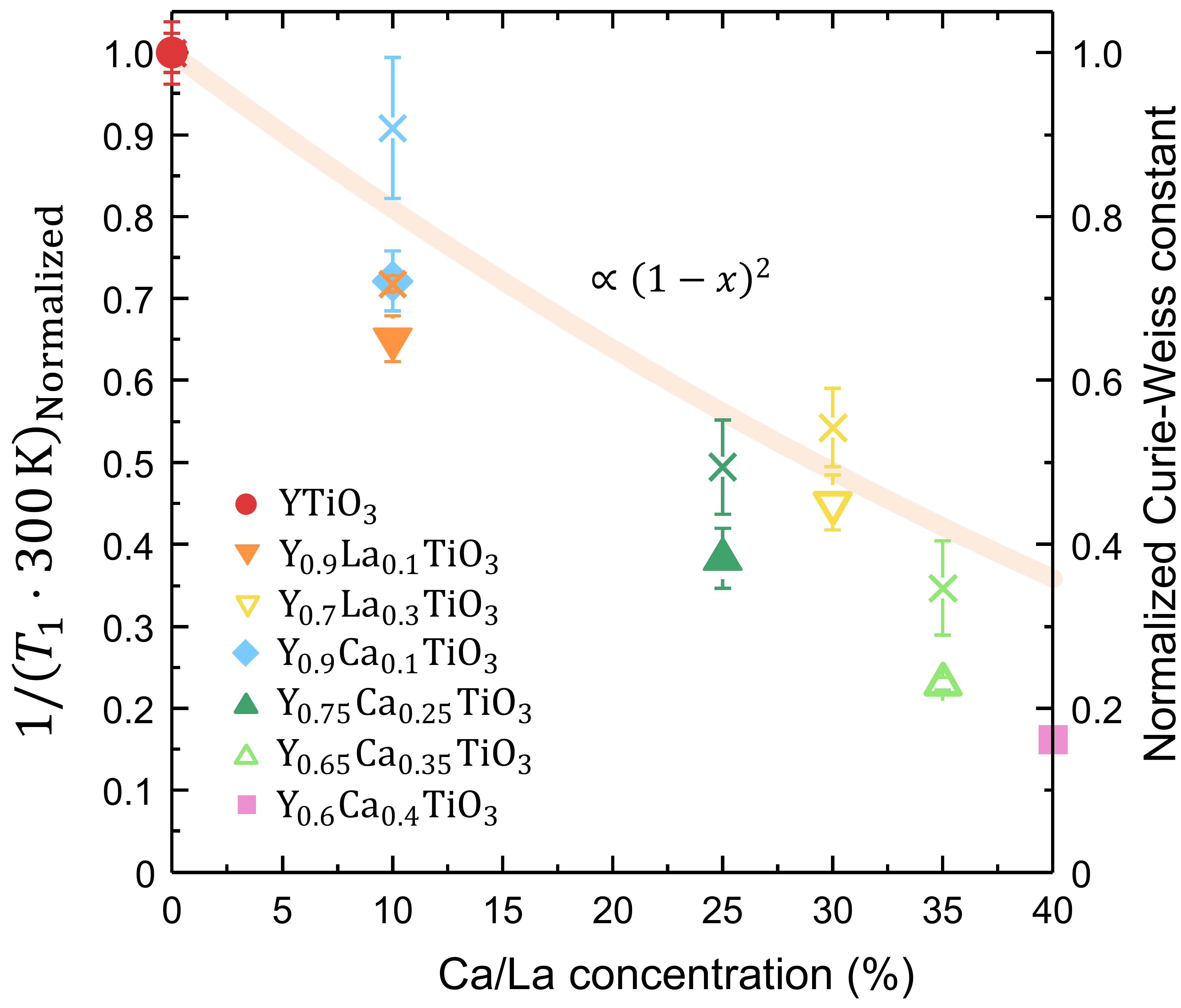}
\caption{\label{fig:T1300} High-temperature spin fluctuations quantified through $1/(T_1T)$ at 300\,K (normalized to the value for YTiO$_3$; full symbols) and Curie constants obtained from Curie-Weiss fits ($\times$ symbols; see Table\,\ref{tab:table1} for fit ranges). The line is the $(1-x)^2$ dependence based on the assumption that substitution/doping effects are captured by simple dilution.}
\end{figure}

In order to obtain insight into the dynamical susceptibility and spin fluctuations, we measured the spin-lattice relaxation time $T_1$. Generally, $1/T_1$ is a measure of the rate of energy exchange between the nuclear spin system and all other components of the material. In our case, the relevant interactions are predominantly between the $^{89}$Y nuclear moments and Ti electronic spins. While it is possible that $T_1$ varies across the NMR line, these measurements were only performed at the frequencies of the NMR signal peaks in Fig.\,\ref{fig:freq} due to the prohibitively long acquisition times that would be needed to determine the full frequency dependence of $T_1$. Importantly, the $^{89}$Y saturation recovery follows a simple exponential time dependence for all samples at all studied temperatures. In dynamically heterogeneous systems, \textit{e.g.}, materials with spin-glass correlations, the saturation recovery profile broadens significantly and is often modeled as a stretched exponential. The lack of such stretching, therefore, suggests the absence of spin-glass physics in the studied doping and temperature range, in agreement with prior work \cite{HameedLa}. 

If the dominant interaction is between nuclear and electronic spins, the inverse product of $T_1$ and temperature, $1/(T_1\,T)$, can be expressed through the imaginary part of the dynamic electronic spin susceptibility \cite{MoriyaT1sus}, $\chi(q,\omega)$:
\begin{equation}
 \frac{1}{T_1 T}=\frac{\gamma_n^2 k_B}{2 \mu_B^2}\sum_{q}|A_q|^2\frac{\textit{Im}\chi(q, \omega_0)}{\omega_0},
\end{equation}
where $\gamma_n$ is the nuclear gyromagnetic ratio, $k_B$ and $\mu_B$ the Boltzmann constant and Bohr magneton, respectively, $\omega_0$ the nuclear Larmor frequency, $q$ are wavevectors in the first Brillouin zone, and $A_q=\sum_i A_i \exp(iqr_i)$, with $A_i$ the hyperfine coupling between the nuclear spin and the electron spin at site $r_i$. As noted, the orthorhombic space group $Pnma$ leads to a peak of the form factor $A_q$ at $q=0$ for the Y site \cite{NMRFuru}. The measured values of $1/(T_1T)$ in dependence on temperature are plotted in Fig.\,\ref{fig:T1}; note that the temperature ranges for some samples are limited due to short spin-spin relaxation times that precluded reliable $T_1$ measurements at low temperatures. Our data reproduce the peak previously seen in YTiO$_3$ \cite{NMRT1}, but at a slightly higher temperature. A possible reason for this difference could be smaller deviations from oxygen stoichiometry in our samples. It is known that deviations from stoichiometry decrease $T_C$ \cite{hameedgrowth}, and the sample we measured had a $\sim 20$\% higher $T_C$ than the one in \cite{NMRT1}. In Fig.\,\ref{fig:T1} the data are compared to a simple Curie-Weiss law, $1/T_1T\propto C_{\textrm{NMR}}/(T-\theta)$, with $\theta$ the Weiss temperature and $C_{\textrm{NMR}}$ the NMR-derived Curie constant. From this comparison, it is clear that the peak results from a decrease of $1/(T_1T)$ below the Curie-Weiss curve. Similar, but less pronounced features are seen in the non-stoichiometric samples, \textit{i.e.}, $1/(T_1T)$ shows a downward deviation from Curie-Weiss behavior below a crossover temperature that depends on the La/Ca concentration. We note that the temperature ranges for some of the samples are limited due to strong increases in spin-spin relaxation rates (see below), which precluded $T_1$ measurements at low temperatures. 

\begin{table}[b]
\caption{\label{tab:table1}%
Weiss constants $\theta$ obtained from the high-temperature Curie-Weiss fits in Fig.\,\ref{fig:T1}, along with the temperature ranges used for the fits. Sample $T_C$ values were obtained from magnetometry \cite{HameedLa,HameedCa, HameedCaErr} except for the case of 25\% Ca doping, where $\mu$SR was used \cite{Ca25}.
}
\begin{ruledtabular}
\begin{tabular}{cccc}
 Sample & $T_C$& Fit range&
$\theta$\\
\colrule
YTiO$_3$ & 30\,K& [140\,K, 300\,K]& $51\pm5$\,K\\
Y$_{0.9}$La$_{0.1}$TiO$_3$ & 17\,K& [100\,K, 300\,K]& $9\pm4$\,K\\
Y$_{0.9}$Ca$_{0.1}$TiO$_3$ &  11\,K& [80\,K, 300\,K]& $19\pm12$\,K\\
Y$_{0.7}$La$_{0.3}$TiO$_3$ & 8.5\,K& [80\,K, 300\,K]&  $3\pm11$\,K\\
Y$_{0.75}$Ca$_{0.25}$TiO$_3$ & 2\,K & [60\,K, 300\,K]& $13\pm15$\,K\\
Y$_{0.65}$Ca$_{0.35}$TiO$_3$ &  PM& [50\,K, 300\,K]& $-27\pm28$\,K\\
\end{tabular}
\end{ruledtabular}
\end{table}

\begin{figure*}
\includegraphics[width=175mm]{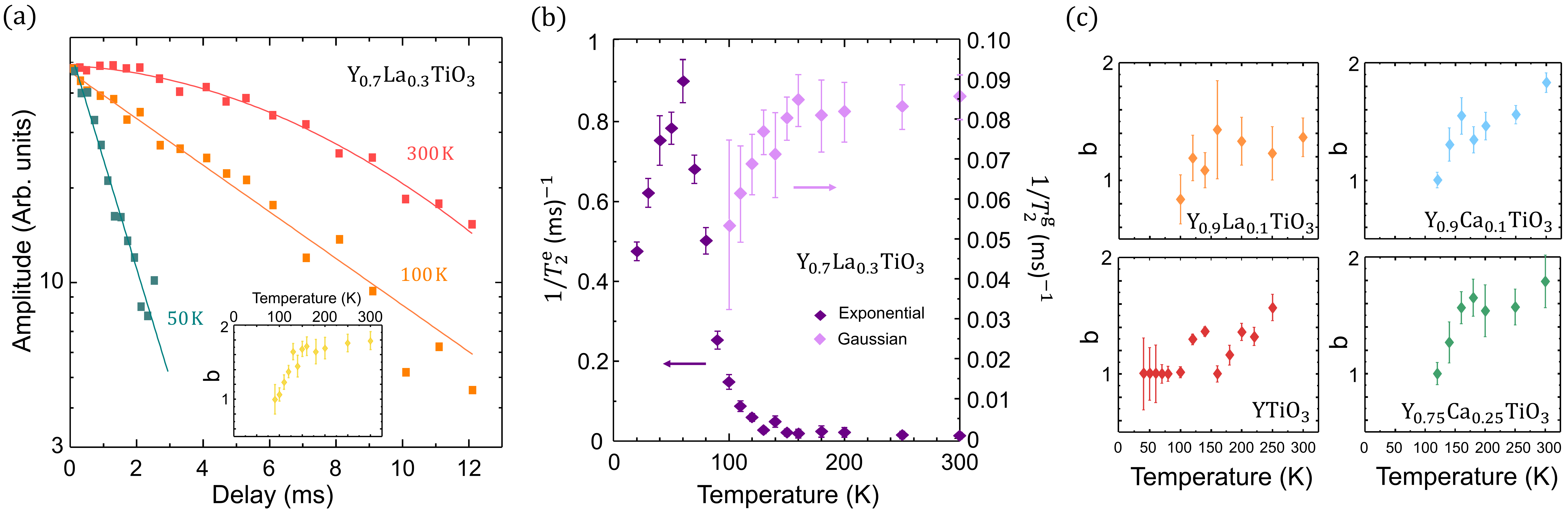}
\caption{\label{fig:La30} (a) Raw spin-echo decay data used to determine $T_2$ relaxation times for Y$_{0.7}$La$_{0.3}$TiO$_3$ at several temperatures. Lines are best fits to the compressed exponential dependence, Eq.\,(3). Insert shows the compression parameter b at different temperatures, demonstrating the transition from nearly Gaussian, $b = 2$, to exponential relaxation, $b = 1$, in cooling. (b) Alternative analysis of the data in (a), using the product of an exponential and Gaussian decay as a fit function. The corresponding values of $T_{2g}$ and $T_{2e}$ are shown, with the left and right axis referring to $1/T_{2e}$ and $1/T_{2g}$, respectively. As in (a), a crossover from Gaussian to exponential relaxation is observed in cooling, with a strong peak in $T_{2e}$. (c) Temperature dependence of the compression parameter $b$, obtained from fits of equation (3) to the raw spin echo decay data in YTiO$_3$ and nonstoichiometric samples. The crossover from Gaussian to exponential echo decay is not as clear as in Y$_{0.7}$La$_{0.3}$TiO$_3$, but a similar trend is seen in all samples.}
\end{figure*}

The Curie-Weiss fits were performed in the high-temperature regime well above $T_C$. The extracted Weiss temperatures $\theta$ roughly match the experimental $T_C$ values (see Table\,I), with the possible exception of YTiO$_3$, where the fit gives $\theta=51$\,K, nearly a factor of two higher than the measured bulk $T_C$ of 30\,K. With increasing La and Ca concentration, the standard deviations and the values of $\theta$ become comparable due to the limited fit ranges, but are still close to the experimental $T_C$ in all cases. At temperatures significantly above $T_C$, the Curie-Weiss expression describes the data well and can be used to investigate the effect of doping/substitution on the mean-field magnetic fluctuations. To that end, we compare in Fig.\,\ref{fig:T1300} $1/(T_1T)$ at $T=300$\,K for all samples with $C_\textrm{NMR}$. Both quantities are normalized to the respective values for YTiO$_3$ and show similar behavior, as expected for $T>>\theta$. 

Interestingly, we do not observe significant differences between Ca doping and La substitution in the high-temperature mean-field regime. In a simplified view, this suggests that doping/substitution amounts to (homogenous) dilution of the ferromagnetism, somewhat analogous to observations for the antiferromagnetic correlations in the electron doped cuprate Nd$_{2-x}$Ce$_x$CuO$_{4\pm\delta}$ \cite{MangPhysRevLett.93.027002}. In Curie-Weiss theory of the macroscopic susceptibility, the Curie constant is proportional to $\mu^2$, where $\mu$ is the \textit{total} size of the individual Ti moments. However, the NMR-derived $C_{\textrm{NMR}}$ includes form-factor effects and is not directly comparable to the results of macroscopic susceptibility measurements \cite{Zhou05}. Eight nearest-neighbor Ti spins add roughly equally to the hyperfine field at the Y site and, as noted, only the ferromagnetic components of the Ti fields contribute. $^{89}$Y NMR thus effectively provides a short-range average of the ferromagnetic component of the Ti spin susceptibility. Therefore, the doping-induced decrease of the effective Curie constant $C_{\textrm{NMR}}$ can either be due to true dilution -- some of the Ti ions converting to a zero-spin state -- or a shift from ferromagnetic to antiferromagnetic fluctuations. The two distinct scenarios are likely related to Ca and La substitution, respectively, but $^{89}$Y NMR cannot distinguish between them. If we assume that the only effect of a nonzero Ca/La concentration is to decrease the FM component of the moment as $1-x$, we obtain $C_{\textrm{NMR}} \propto (1-x)^2$. This is in reasonably good agreement with the data (Fig.\,\ref{fig:T1300}), which suggests that, at least far from the magnetic ordering temperatures, the spin system is essentially homogeneous on the nanoscale and insensitive to the details of the electronic interactions. However, the observed decrease of the dynamical susceptibility below the Curie-Weiss dependence is unexpected for materials with FM order, where the $q=0$ susceptibility generically diverges at the magnetic transition. This conundrum is further deepened by the fact that the static susceptibility, as seen from the NMR shifts, shows the expected increase near $T_C$, especially in YTiO$_3$. We relate this to our ESR results and discuss possible explanations in more detail below.

\begin{figure}
\includegraphics[width=0.4\textwidth]{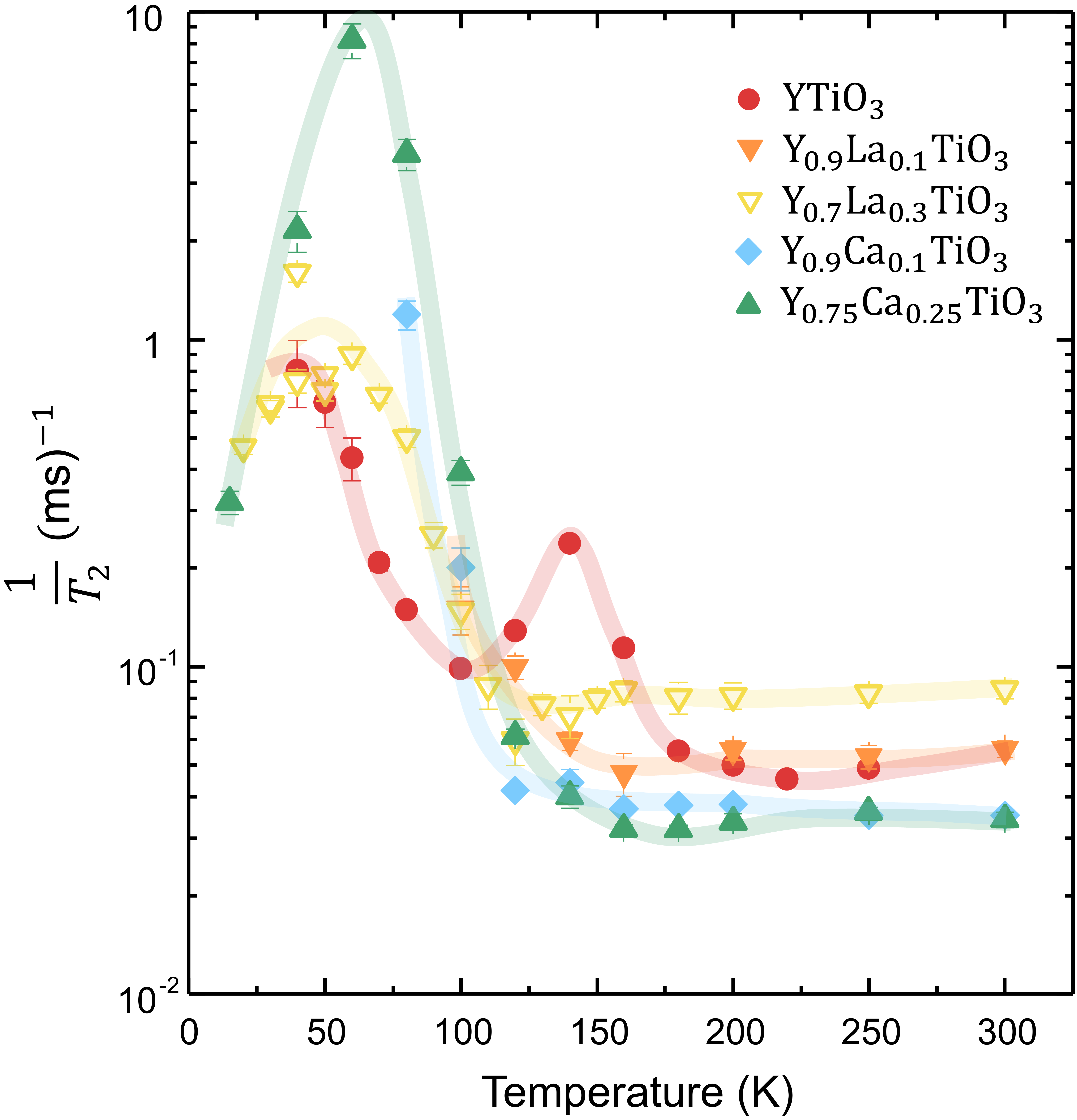}
\caption{\label{fig:T2} $1/T_2$ values obtained from fits to a compressed exponential form, Eq.\,(3). Peaks are observed close to the temperatures where $1/T_1 T$ starts to deviate from Curie-Weiss behavior, and the peak heights increase with increasing Ca/La concentrations. The low-temperature increase in YTiO$_3$ is likely due to fluctuations in the vicinity of $T_C = 30$\,K, whereas for other samples the peak might be too close to $T_C$ for the additional increase to be resolved. Lines are guides to the eye. }
\end{figure}

Nontrivial temperature dependences are also observed for the spin-spin relaxation time, $T_2$. The inverse $1/T_2$ is a measure of the nuclear spin decoherence rate; in contrast to $T_1$, $T_2$ is often not easy to describe theoretically, but it can be very sensitive to spin fluctuations. Qualitatively, we observe a change in the character of the spin-spin relaxation, from nearly Gaussian at high temperatures to exponential below $\sim 100$\,K. Representative plots of this behavior in Y$_{0.7}$La$_{0.3}$TiO$_3$ are shown in Fig.\,\ref{fig:La30}\,(a) and \ref{fig:La30}\,(b). These can be phenomenologically analyzed in two previously established ways \cite{JulienNature,Walstedt}. First, we treat the spin echo decay function as a product of Gaussian and exponential components,
\begin{equation}
    Amplitude \propto e^{-(\frac{\tau}{T_{2,e}})}\cdot e^{-(\frac{\tau}{T_{2,g}})^2},
\end{equation}
where $\tau$ is the echo delay, and $T_{2,e}$ and $T_{2,g}$ are the exponential and Gaussian relaxation times, respectively. The data for Y$_{0.7}$La$_{0.3}$TiO$_3$ (Fig.\,\ref{fig:La30}\,(a)) most clearly show the crossover from Gaussian to exponential relaxation with decreasing temperature, and similar behavior is seen in all samples. However, Eq.\,(2) is not well suited for limiting cases of purely exponential or Gaussian relaxation. To remedy this, and obtain a single $T_2$ parameter for a comparison of data at different temperatures and among all samples, we use a second approach -- a fit to a compressed exponential echo decay of the form 
\begin{equation}
    Amplitude \propto e^{-(\frac{\tau}{T_2})^b},
\end{equation}
where $1<b<2$. In the limiting cases of $b = 1$ and $b = 2$, pure exponential and Gaussian relaxation are recovered, respectively, and the parameter $b$ can be used as a convenient diagnostic of the functional form of the relaxation process. The fitted values of $b$ are shown in Fig.\,\ref{fig:La30}, with the corresponding $T_2$ values plotted in Fig.\,\ref{fig:T2}. These results indicate the presence of a significant peak in the (exponential) spin-spin relaxation rate, roughly at the temperatures where $1/(T_1T)$ starts to deviate from Curie-Weiss behavior. In some samples, the peak is so strong that $T_2$ becomes too short for spin-echo measurements. 

As noted, $T_2$ is not easy to understand quantitatively, but some inferences can be made. An upper limit to the relaxation time is provided by the dipole-dipole coupling between the $^{89}$Y nuclear spins, which is known to lead to Gaussian spin-spin relaxation if mutual spin flips are negligible. In this case, the broadening of magnetic resonance lines due to direct dipole-dipole interactions is well described by van Vleck theory \cite{vanVleckT2}. The width of the resonance line is given by
\begin{equation}
    \big<\Delta\nu^2\big>_{Av}=30\,\Big(\dfrac{\mu_0}{4\pi}\Big)^2g_n^4 \mu_n^4 h^{-2} d^{-6} \Bigg[\dfrac{1}{3}I(I+1)\Bigg],
\end{equation}
where $g_n$ and $\mu_n$ are the nuclear g-factor and magneton, which are proportional to the known reduced gyromagnetic ratio of $^{89}$Y, $\frac{\gamma}{2\pi}=$\,2.08637\,MHz/T. $^{89}$Y is a spin $I=1/2$ nucleus with 100\% natural abundance, and the distance between neighboring sites is $d\approx 3.8$\,\AA. This information allows us to approximate the spin-spin relaxation time simply as $T_2=\dfrac{1}{\Delta\nu}$, which results in the estimate $T_{2,g}^{dip}\approx$\,70\,ms. The experimental $T_{2,g}$ values at high temperatures are systematically lower than this limit, by a factor 2-5, depending on the Ca/La concentration. The Y-Y dipolar coupling is therefore enhanced compared to the simple direct interaction estimate and might involve, \textit{e.g.}, indirect exchange coupling \cite{Walstedt} or relaxation through fluctuations of the titanium nuclear spins \cite{T2modeling}. 

\begin{figure*}
\includegraphics[width=175mm]{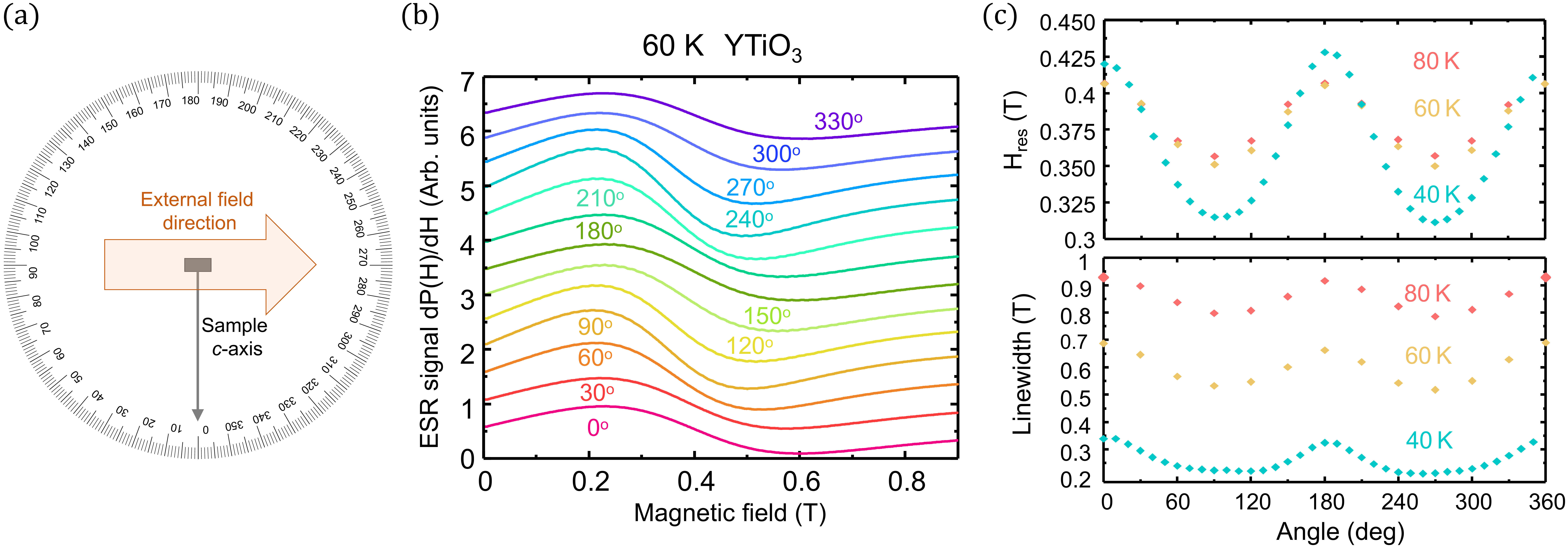}
\caption{\label{fig:esr1} (a) Sketch of the magnetic field direction and sample orientation relative to the goniometer angle scale. (b) Magnetic field derivative of the absorbed microwave power at 60\,K for the YTiO$_3$ sample. Results obtained for different orientations are shifted upwards for easier comparison. (c) Angular dependence of the resonant field and width of the ESR absorption line for YTiO$_3$ at 40\,K, 60\,K and 80\,K obtained from Lorentizan fits.}
\end{figure*}

A crossover to an exponential spin-spin relaxation regime at low temperatures is also seen in other materials, including cuprate superconductors \cite{cupraterelax,T2modeling,Walstedt}, and appears when nuclear spin-flip processes become important. The spin-spin relaxation originates from fluctuations of the hyperfine field at the Y sites, which might be due to both nuclear and electronic spin fluctuations. In the context of cuprate superconductors, numerical studies of the nuclear spin Hamiltonian have reproduced the crossover when taking into account nuclear unlike-spin coupling and $T_1$ effects \cite{T2modeling}. The $^{89}$Y spins couple to $^{47,49}$Ti spins, whose spin-lattice relaxation can modulate the local fields and affect the Y decoherence process. However, the natural abundance of $^{47,49}$Ti is low, and the calculated exponential decay for the unlike-spin case is, in fact, slower than the Gaussian, since it effectively leads to motional narrowing \cite{T2modeling}. A more plausible possibility is therefore a coupling to the electronic spin subsystem, where an exponential spin-echo decay also appears in the limit when the electronic fluctuations are much faster than the Larmor frequency \cite{JulienNature,Takigawa86}. The spin-spin relaxation rate is then proportional to the electronic spin correlation time. In this scenario, the peaks in Fig.\,\ref{fig:T2} signify that a significant slowing of the electronic dynamics takes place around the characteristic crossover temperatures. This is more clearly visible in $1/T_2$ than $1/(T_1T)$ due to the fact that the spin-lattice relaxation only depends on the electronic susceptibility at the Larmor frequency, whereas spin decoherence can be affected by fluctuations at far shorter timescales, where the changes are more dramatic.

Our study includes one sample that lies close to the insulator-metal boundary, Y$_{0.65}$Ca$_{0.35}$TiO$_3$, and one sample with a temperature-induced insulator-metal transition, Y$_{0.6}$Ca$_{0.4}$TiO$_3$. The latter composition is metallic below $\sim 120$\,K, and the NMR signal becomes extremely weak in the metallic phase, most likely due to the small radiofrequency penetration depth. A recent x-ray absorption spectroscopy study uncovered strong evidence for phase separation associated with the insulator-metal transition \cite{HameedCa, HameedCaErr}. However, we did not observe any indication of phase separation in NMR: no secondary resonance lines are seen, and $T_1$ remains single-valued in the insulating state close to the transition. This does not necessarily exclude a phase separation scenario, since it is possible that the signal from nanoscale metallic regions is significantly shifted and/or weakened. More work is needed to definitively confirm or refute the existence of microscopic phase separation on the NMR timescale.

\section{ESR results}

\begin{figure}
\includegraphics[width=0.45\textwidth]{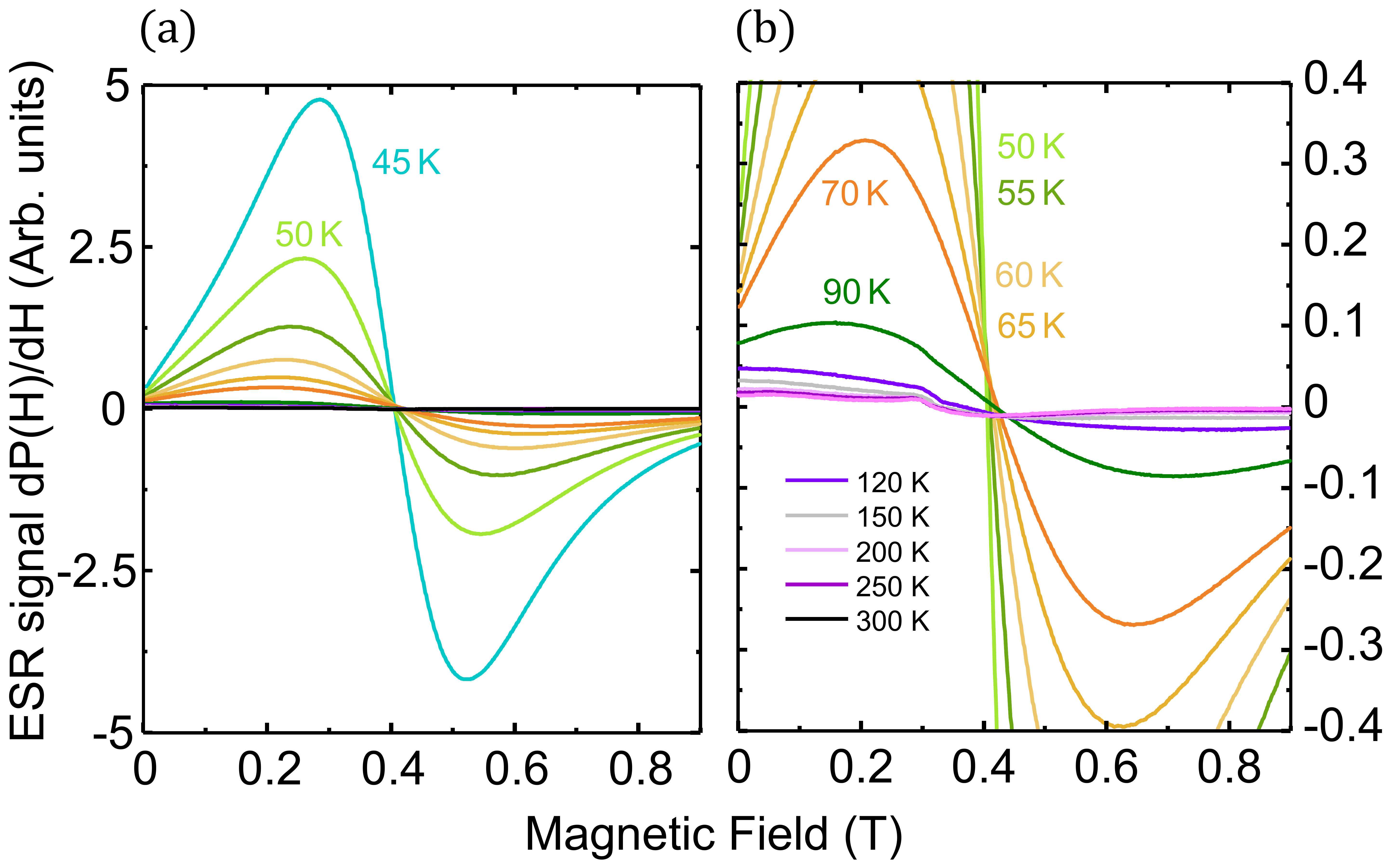}
\caption{\label{fig:yto_ESR_000} Magnetic field derivative of the absorbed microwave power measured for YTiO$_3$ at different temperatures, with the external field perpendicular to the crystallographic $c$-axis. A strong increase of the signal is observed with decreasing temperature, as expected for a system with FM spin fluctuations. We show measurements at all temperatures in (a), whereas (b) shows spectra at higher temperatures with the vertical scale adjusted for clarity.} 
\end{figure}

ESR spectra were measured as a function of temperature and sample orientation with respect to the external magnetic field. The orientation is shown schematically in Fig.\,\ref{fig:esr1}\,(a). At 0$^\circ$ and 180$^\circ$, the crystallographic $c$-axis is oriented perpendicular to the field, and at 90$^\circ$ and 270$^\circ$ it is parallel to the field. The resonant field and absorption linewidth both depend on the sample orientation, as can be seen from the raw spectra for YTiO$_3$ at 60\,K (Fig.\,\ref{fig:esr1}\,(b)). This field orientation dependence can be seen well above $T_C$ (Fig.\,\ref{fig:esr1}\,(c)), which is not unexpected given the low structural symmetry and known magnetic anisotropy. Figure\,\ref{fig:yto_ESR_000} shows the raw ESR absorption lines for YTiO$_3$ up to room temperature, with the sample $c$-axis perpendicular to the external magnetic field. The signal significantly decreases with increasing temperature. This is expected due to the decrease of electronic spin susceptibility and in agreement with the NMR shift results.

\begin{figure}
\includegraphics[width=0.45\textwidth]{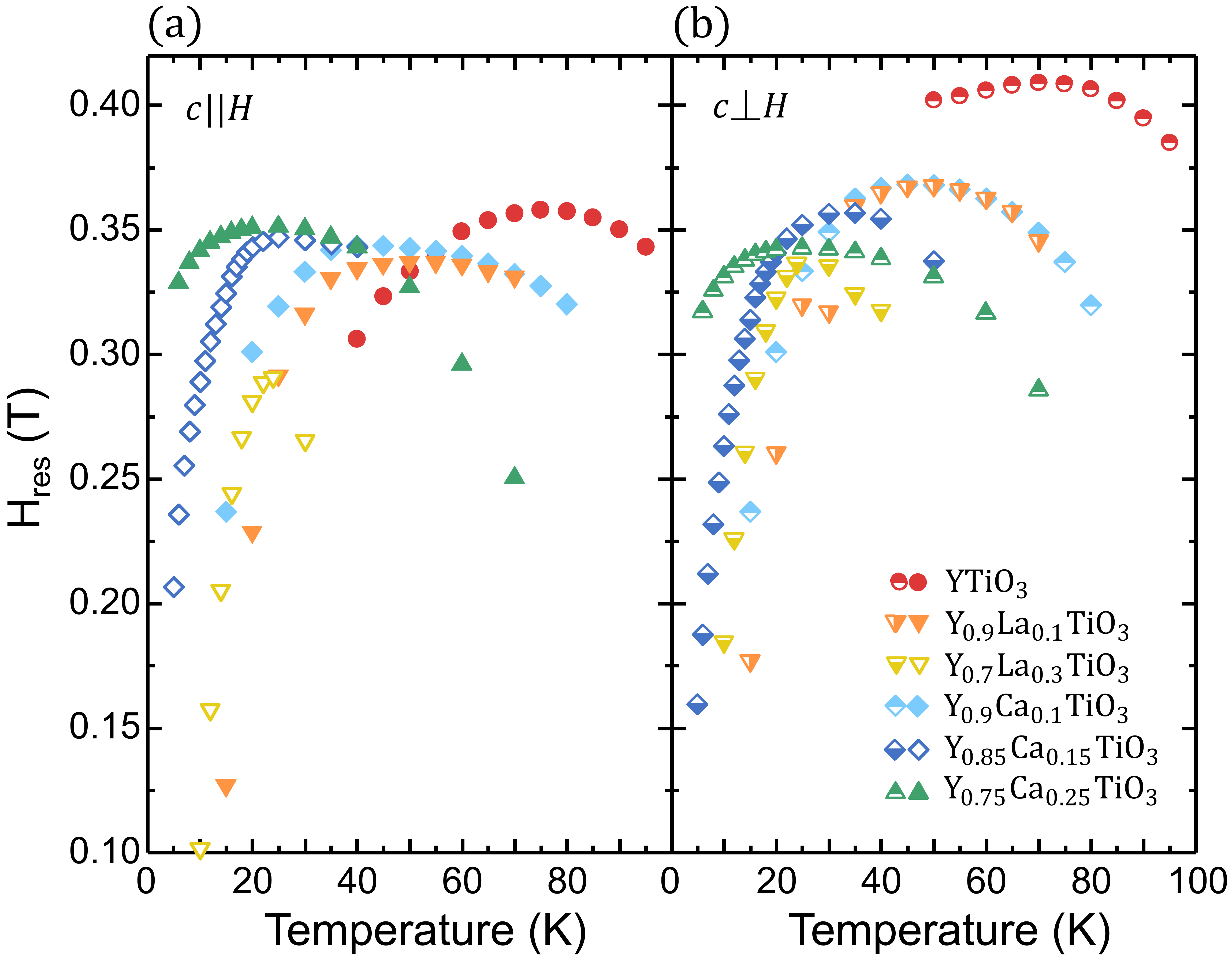}
\caption{\label{fig:Hres} Resonance field $H_{res}$ of the ESR signal obtained from Lorentzian fits. Temperature and doping dependence of $H_{res}$ with the external field:  (a) parallel to the $c$-axis; (b) perpendicular to the $c$-axis.} 
\end{figure}

\begin{figure}
\includegraphics[width=0.45\textwidth]{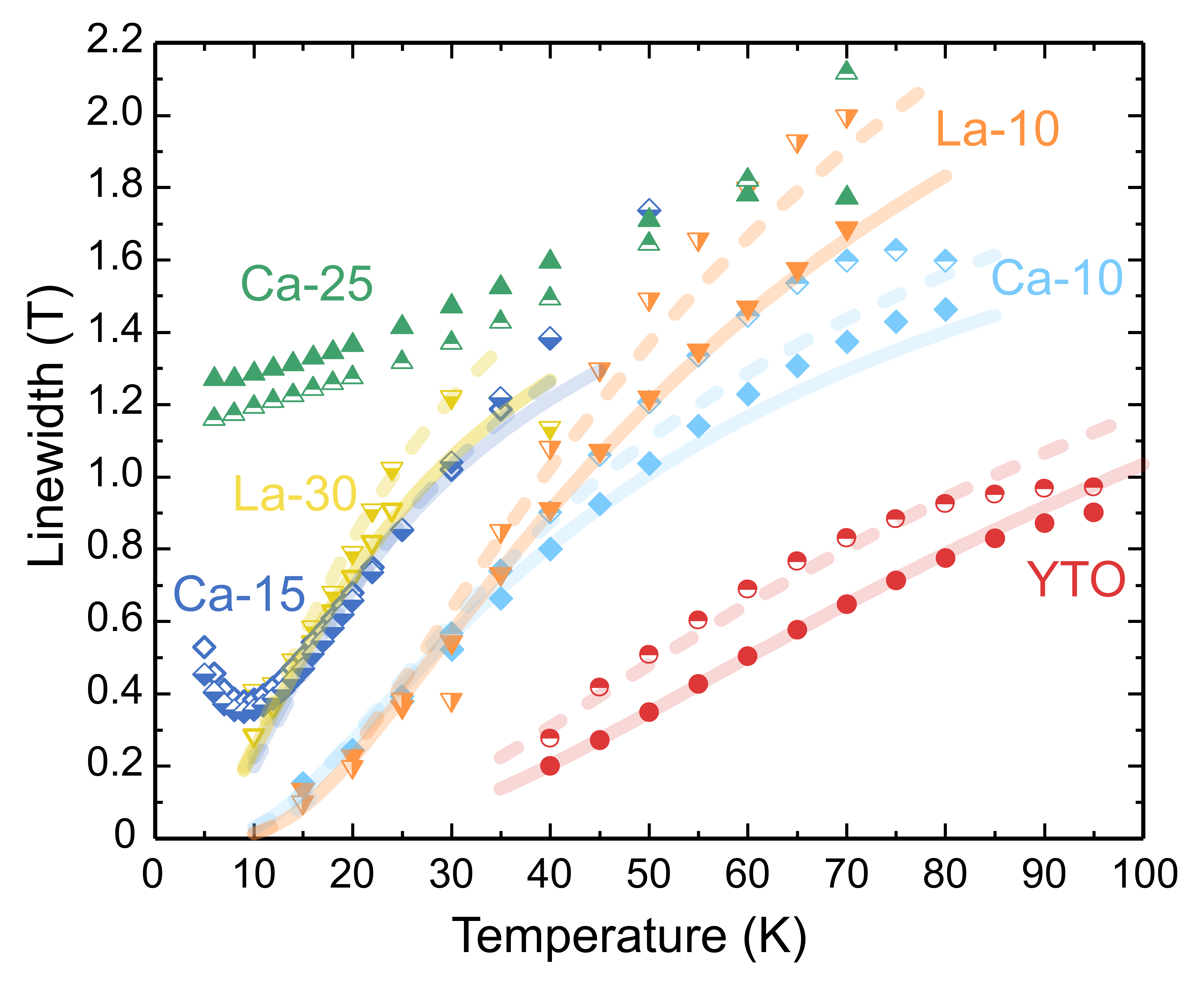}
\caption{\label{fig:lw} Lorentzian linewidths of the ESR absorption lines vs. temperature, with the $c$-axis parallel (full symbols) and perpendicular (half-empty symbols) to the external field. Error bars from fits are smaller than the symbol size, but we note that systematic errors are possible when the widths become comparable to the experimental magnetic field of $\sim 1$\,T. The linewidths are significantly broader than the static dipole-dipole estimate (see text), and are instead determined by fast spin-lattice relaxation. Lines are best fits to Eq.\,(5).} 
\end{figure}

\begin{figure}
\includegraphics[width=0.48\textwidth]{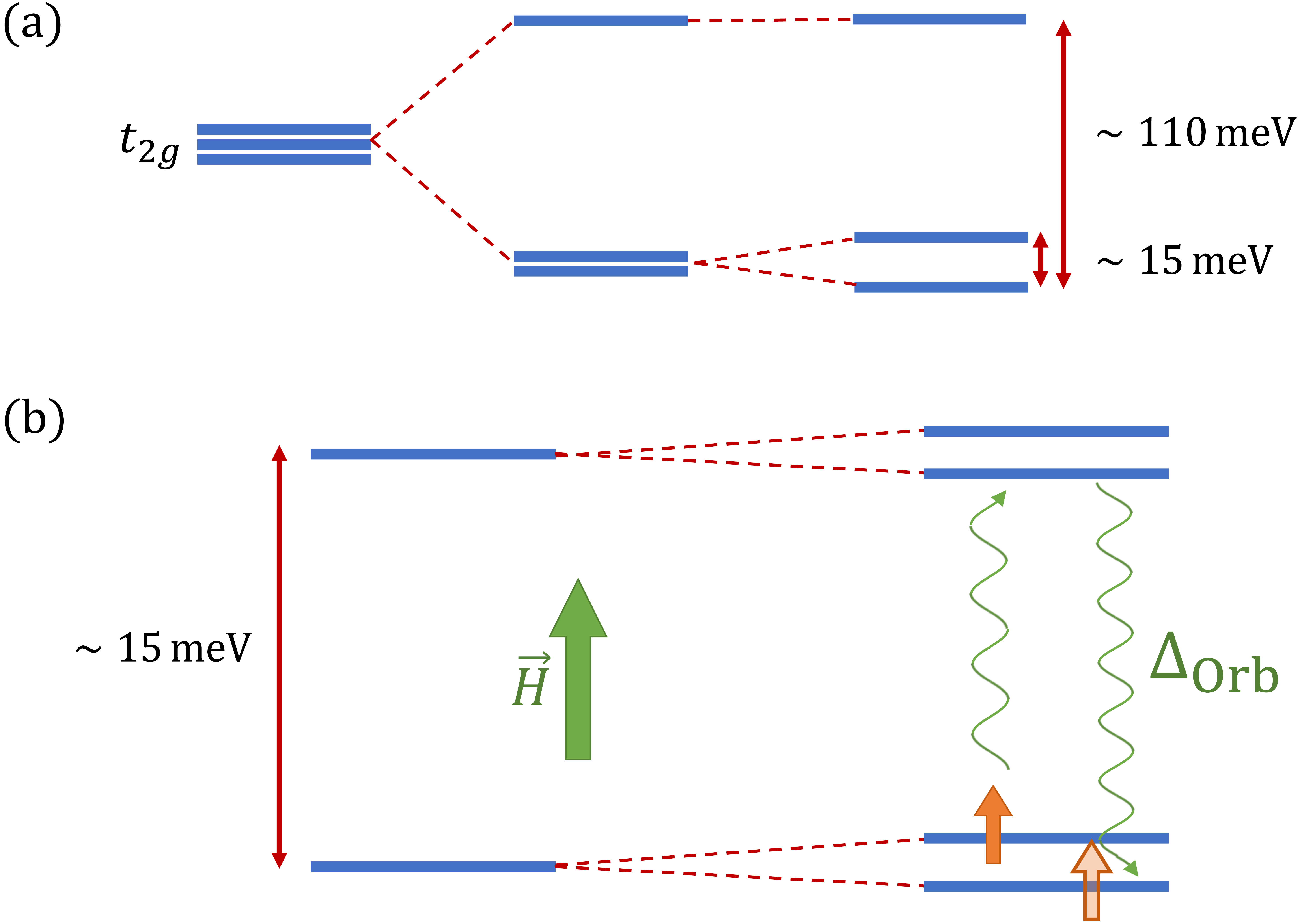}
\caption{\label{fig:t2g} (a) Schematic depiction of the splitting of the triply-degenerate crystal field levels by Jahn-Teller distortions of the TiO$_6$ octahedra. Octahedral elongation along the $c$-axis creates a doubly-degenerate ground state, while additional distortions in the basal plane lift the orbital degeneracy completely. Energy splitting values were taken from \cite{revijalac}. (b) Schematic of the resonant phonon (Orbach) relaxation process of the electronic spins, that involves a real electronic excited state spin doublet separated by the ground state spin doublet by $\Delta_{Orb}$. The external magnetic field $H$ lifts the remaining two-fold spin degeneracy of the orbital levels. Note that this Zeeman splitting is much smaller than $\Delta_{Orb}$ in our case.} 
\end{figure}

The temperature dependence of the absorption signal above $T_C$ was measured for all samples in two orientations, \textit{i.e.}, with the $c$-axis roughly parallel and perpendicular to the external magnetic field. All lineshapes were well fit by a Lorentzian lineshape derivative, from which we calculated the resonant field $H_{res}$ and linewidth of the ESR signal. Resonant field values for the two sample orientations are shown in Fig.\,\ref{fig:Hres}. In YTiO$_3$, on approaching $T_C$, significant sample magnetization creates magnetic domains, which generate artifacts that distort the signal. The measurements were therefore conducted above 40\,K. The largest anisotropy of $H_{res}$ for different orientations was seen in YTiO$_3$ (Fig.\,\ref{fig:esr1}\,(c)), consistent with magnetic anisotropy observed previously in the ordered state \cite{HameedLa}. All samples show a significant shift of the resonant field to lower values as $T_C$ is approached. The onset temperature of this shift is 10-20\,K above the FM transition. At higher temperatures the resonant field approximately saturates, whereas the linewidth continues to increase (Fig.\,\ref{fig:lw}). The latter is the most interesting feature of the ESR results, and we analyze it in detail in what follows.

The ESR linewidth is most commonly determined by electronic spin-spin interactions, with two distinct regimes: dipolar broadening and exchange narrowing. In the former, dipole-dipole interactions between spins that are quasistatic on the ESR timescale broaden the line and lead to a Gaussian lineshape; this is the ESR equivalent of the $^{89}$Y NMR estimate of $T_{2,g}$ discussed in Section III. Conversely, in the exchange narrowing regime, the spins fluctuate fast enough to effectively average out the local fields, which leads to long coherence times and exponential spin-spin relaxation, \textit{i.e.}, a narrow Lorentzian lineshape. Similar to the calculation of NMR $T_{2,g}$, we can use van Vleck theory to estimate the Gaussian linewidth in the dipolar broadening regime, which provides an upper limit for the linewidth from spin-spin interactions. Applying Eq.\,(4) to the Ti ion, we take $d=3.9\,$\AA\ to be the distance between neighboring ions, along with values of the free electron $g$-factor and Bohr magneton. Frequency-to-field conversion \cite{MabbsESR} gives an estimated linewidth of around 0.09\,T. Again, this is an upper limit, and the likely presence of exchange narrowing would further decrease the spin-spin linewidth. As can be seen from Fig.\,\ref{fig:lw}, the absorption lines in all samples are broader than the static dipolar limit, and the lineshapes are consistent with a Lorentzian, rather than Gaussian functional form. Therefore, the linewidths are not determined by spin-spin relaxation processes in the entire temperature range of interest, and instead must be due to spin-lattice relaxation.

\begin{figure}
\includegraphics[width=0.45\textwidth]{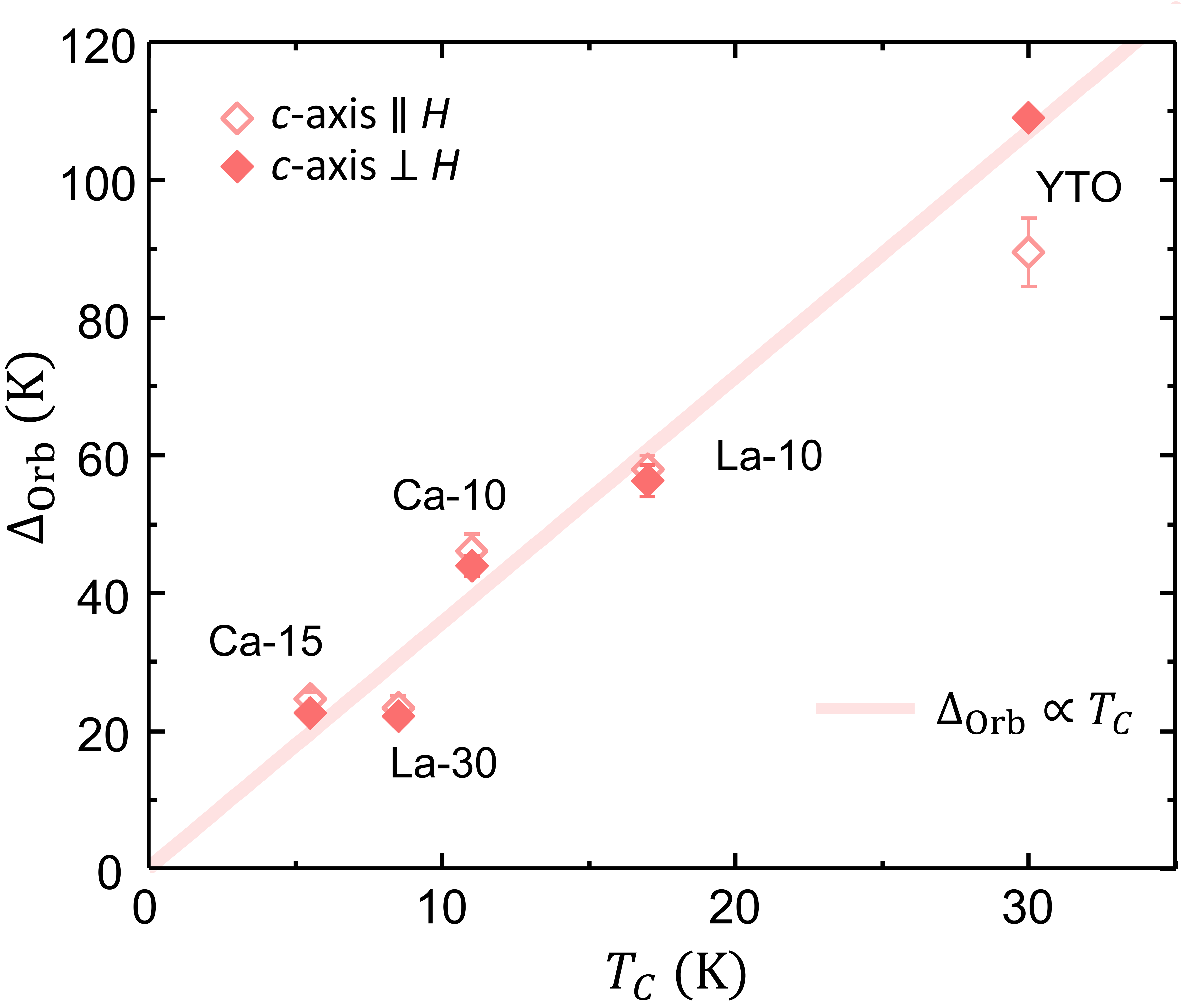}
\caption{\label{fig:Orb} Values of the $\Delta_{\text{Orb}}$ gap in Kelvin, obtained from fits of the Orbach relaxation temperature dependence to ESR linewidths. The line is a linear fit to the data with the magnetic field parallel to the crystallographic c-axis (full symbols).} 
\end{figure}

Short $T_1$ relaxation times that homogeneously broaden ESR lines are a known feature in transition metal oxides \cite{AbragamESR, wileyepr}. Three common relaxation processes lead to distinct temperature dependencies of $1/T_1$. The direct phonon process, which involves the exchange of one phonon per relaxation event and only becomes important at very low temperatures, would produce a linear temperature dependence, but likely does not apply to our case. The Raman phonon process, which involves the exchange of two phonons of different frequencies (with a significantly larger phase space than the direct process), generates a distinct $1/T_1 \propto T^9$ temperature dependence. This is clearly not in agreement with the experimental results. Lastly, the acoustic phonon Orbach process yields an activated exponential temperature dependence \cite{Orbach,HuangPhysRev.154.215}, which describes the data reasonably well. Relaxation through the Orbach process proceeds in two steps (see Fig.\,\ref{fig:t2g}\,(b)): first, a phonon is absorbed to transfer the electronic system from the upper Zeeman level to a real electronic excited state that is separated from the ground state Zeeman doublet by a gap $\Delta_{Orb}$. A phonon is then emitted, and the electronic system relaxes to the lower Zeeman level. This is therefore a resonant two-phonon relaxation process. If $\Delta_{Orb}$ is significantly larger than the Zeeman splitting, but still within the acoustic phonon spectrum, the temperature dependence of the linewidth becomes \cite{Orbach}:
\begin{equation}
    \Delta H\propto \dfrac{1}{T_1} \propto \exp(-\dfrac{\Delta_{\text{Orb}}}{k_BT}).
\end{equation}
In principle, the linewidth contains an additional constant term that takes into account the dipolar effects and inhomogeneous broadening. Yet this constant is sufficiently small to be negligible in the relevant temperature range, and we have not included it in our fits in order to minimize the number of free parameters. This assumption is also supported by the nearly perfectly Lorentzian lineshapes, the small estimated upper limit for dipolar broadening, and the good quality of two-parameter fits to Eq.\,(5) (Fig.\,\ref{fig:lw}). $\Delta_{\text{Orb}}$ values obtained by fitting Eq.\,(5) to the measured linewidths in Fig.\,\ref{fig:lw} are expressed in Kelvin and plotted in Fig.\,\ref{fig:Orb} vs. the zero-field $T_C$ values. We find a fairly accurate linear relationship between the two, with a slope of $3.6\pm 0.2$. Moreover, the gaps obtained for the two magnetic field orientations are quite similar, despite the linewidth anisotropy.

We note that an optical phonon Orbach process could also be a possible candidate for the relaxation mechanism. In this case, $\Delta_{\text{Orb}}$ corresponds to an optical phonon frequency. However, there are no zone-center phonon modes in the RE titanates with energies that match the fitted values of $\Delta_{\text{Orb}}$ or their strong doping dependence \cite{Kovaleva3, Sugai}. Therefore, we believe that $\Delta_{\text{Orb}}$ must correspond to an electronic excited state. As in similar transition metal oxides with octahedral coordination, the Ti atomic $d$ orbitals transform to $t_{2g}$ and $e_g$ orbitals. In RE titanates, the triply degenerate $t_{2g}$ levels may be split further by interactions in the material. Jahn-Teller distortions along the crystallographic $c$-axis partially lift the $t_{2g}$ degeneracy. In YTiO$_3$, this pushes the excited state $100-200$\,meV above the ground state energy \cite{revijalac,disa,streltsovCF}. However, a further lifting of the degeneracy of the lowest $t_{2g}$ orbitals can occur through in-plane Jahn-Teller and/or spin-orbit coupling effects (Fig.\,\ref{fig:t2g}), and the observed $\Delta_{\text{Orb}}$ most likely corresponds to this splitting. In previous work, an analysis of the Coulomb repulsion between the electron in the Ti $t_{2g}$ orbital and the ligand O$^{2-}$ ion within the point charge approximation provides an estimate of $\sim 15$\,meV for Jahn-Teller-induced splitting between the lowest $t_{2g}$ levels in YTiO$_3$ \cite{mociImada,revijalac}. In our experiment, we obtain $\Delta_{\text{Orb}} \sim 10$\,meV $\sim 100$\,K, which is in fairly good agreement the calculations. It is also expected that the Jahn-Teller distortion decreases with Ca/La substitution \cite{revijalac}, and it has been suggested that the antiferro-orbital arrangement in the basal plane promotes FM interactions between the Ti spins \cite{revijalac}. This agrees well with the experimentally observed correlation between $\Delta_{\text{Orb}}$ and $T_C$. We also note that a similar analysis of ESR linewidths has previously uncovered a splitting of the $t_{1u}$ orbital in the C$_{60}^-$ anion by the Jahn-Teller distortion \cite{C60}.

\section{Discussion and summary}

The most important result of our study is the identification of the gap scale $\Delta_{\text{Orb}}$, which is essential to understand the electronic structure and low-energy orbital physics of Y(Ca,La)TiO$_3$. We believe that most of the unconventional magnetic fluctuation effects above $T_C$ observed here and in previous work \cite{knafo,kovaleva1,cheng} can be traced to the existence of this orbital splitting. It is likely not a coincidence that phonon shifts and metastable ferromagnetism appear at temperatures comparable to $\Delta_{\text{Orb}}$ in YTiO$_3$. Moreover, the appearance of a nonzero $\Delta_{\text{Orb}}$ in FM samples and its clear correlation with $T_C$ (Fig.\,\ref{fig:Orb}) indicates a relation between the complete lifting of orbital degeneracy and ferromagnetism in RE titanates, as previously suggested \cite{revijalac,sol}. Conversely, in the paramagnetic (and possibly AFM) state, the $t_{2g}$ orbitals are at least doubly degenerate, which implies strong orbital fluctuations. This provides crucial experimental input to understand the unusual ferromagnetism of the RE titanates, and it confirms the notion that FM spin-spin interactions are stabilized by the lifting of orbital degeneracy and accompanying orbital order patterns. 

It is possible, at least qualitatively, to relate the existence of the relatively low energy scale $\Delta_{\text{Orb}}$ to the features we observe in NMR. First, the downturn in the dynamic susceptibility $\chi''(\omega_L) \propto 1/(T_1 T)$ appears at temperatures roughly consistent with $\Delta_{\text{Orb}}$, which indicates that the spin fluctuation spectrum changes considerably when the orbital degeneracy is effectively lifted. However, the static susceptibility $\chi'(0)$, as measured by the NMR shift, continues to increase as $T_C$ is approached. As noted, this is an unusual result that requires further scrutiny. One possible way to explain the disparate temperature dependences of the two susceptibility components is that low-frequency magnetic fluctuations at temperatures below $\Delta_{\text{Orb}}$ become suppressed, with a temperature-dependent spectral weight rearrangement. The components of the susceptibility are not independent, but related through Kramers-Kronig relations, specifically
\begin{equation}
    \chi'(0) = \int_0^{\infty}d\omega \chi''(\omega)/\omega.
\end{equation}
It is then possible to obtain different temperature dependences for $\chi''(\omega_L)$ and $\chi'(0)$ if the spectral weight around $\omega_L$ is transferred to energies either above or below $\omega_L$ (or both). Given that NMR does not provide the full $\chi''(\omega)$, it is not straightforward to deduce the nature of the spectral weight transfer. Fortunately, the ESR results provide additional constraints. In case of a sizable increase of the spectral weight below $\omega_L$, the spin fluctuations would be slower than the ESR timescale, and would thus lead to a shift of the ESR line. Such a shift is indeed observed, albeit somewhat closer to $T_C$ than the scale $\Delta_{\text{Orb}}$. Yet this suggests that quasistatic fluctuations might be present above $T_C$. This is consistent with evidence from $\mu$SR that the FM transition in RE titanates is, in fact, first order \cite{HameedLa}. However, the shifts are much smaller than what would correspond to fully ordered moments, which implies a relatively small quasistatic spectral weight. It is therefore likely that a transfer to higher energies occurs as well, and it is physically reasonable that strong fluctuations only appear above the gap scale $\Delta_{\text{Orb}}$. It would be of interest to determine the complete $\chi''(\omega)$, \textit{e.g.}, via inelastic neutron scattering, to fully resolve this question and obtain further insight into the interplay of orbital and spin fluctuations. If quasistatic fluctuations are indeed present, they might also be observable in sub-MHz AC susceptibility measurements. 

In non-stoichiometric compounds, the Jahn-Teller distortions are likely inhomogeneous in real space due to the different sizes and electronic configurations of Y and Ca/La ions, and a spatial distribution of local level splitting would be expected. For simplicity, we do not include this effect in our ESR analysis, but it might be important to attain a quantitative understanding of the NMR linewidths if the linear relation between $\Delta_{\text{Orb}}$ and $T_C$ (or, equivalently, the local moments) remains valid at the nanoscale. Moreover, we assume a temperature-independent $\Delta_{\text{Orb}}$, but it is conceivable that the gap opens at a characteristic temperature, which would amount to an orbital ordering transition. The relatively sharp peaks observed in $1/T_2$ might indirectly support such a scenario, but since there is no evidence of an abrupt transition from thermodynamic and structural probes \cite{magnetoelast, knafo}, we believe that a crossover is more likely. 

Our results showcase the intricate interplay among structural distortions, orbital and spin degrees of freedom in the RE titanates, which may be relevant for a wide range of complex oxides such as vanadates \cite{koborinai2016} and manganites \cite{dagotto2005}. Perhaps the most interesting implication of our results is that RE titanates without FM order show partial orbital degeneracy, \textit{i.e.}, the long-debated orbital-liquid state \cite{orbliqKeimer,orbliqKhal,revijalac,Zhaoorbliq}. The significant slowdown of the spin fluctuations is linked to the decrease and disappearance of the orbital gap, as indicated by the enhanced $1/T_2$ in Ca/La substituted compounds. A more detailed investigation of the FM-paramagnetic boundary in the Ca-doped system might be highly interesting, in addition to a further exploration of the role of orbital fluctuations in the metallic state of RE titanates. 

In summary, the combination of NMR and ESR has enabled a detailed study of the low-energy magnetic fluctuations in Y(Ca,La)TiO$_3$. Our results demonstrate the unusual nature of the fluctuations and establish a link to orbital physics. ESR in particular has provided interesting and physically transparent insights, most importantly the identification of the electronic gap scale $\Delta_{\text{Orb}}$, due to the fact that the linewidths are dominated by spin-lattice relaxation. The unconventional features observed in NMR -- different behavior of the static and dynamic susceptibilities, and strong peaks in the temperature dependence of $T_2$ -- can be qualitatively connected to the appearance of the small electronic scale $\Delta_{\text{Orb}}$. A quantitative understanding of the NMR results is beyond the scope of the present work and will likely require a microscopic theory and more complete information on the dynamic spin susceptibility.

\acknowledgements

We thank Ivan Jakovac, Marina Ilakovac Kveder, Dijana Žilić, Christoph Wellm, Rafael Fernandes and Zhentao Wang for discussions and comments. The work at the University of Zagreb was supported by the Croatian Science Foundation through Grant No. IP-01-2018-2970, using equipment funded in part through project CeNIKS co-financed by the Croatian Government and the European Union through the European Regional Development Fund - Competitiveness and Cohesion Operational Programme (Grant No. KK.01.1.1.02.0013). The work at the Leibniz Institute for Solid State and Materials Research was funded through the Research Group Linkage Programme of the Alexander von Humboldt Foundation (ref. no. 3.4-1022249-HRV-IP) and by the Deutsche Forschungsgemeinschaft (DFG) through Grant. No. KA1694/12-1. The work at the University of Minnesota was funded by the US Department of Energy through the University of Minnesota Center for Quantum Materials, under Grant No. DE-SC-0016371.


\newpage
\bibliography{main}
\end{document}